\documentclass[11pt,a4paper]{article}

\usepackage{bm, amssymb, amsmath, graphicx, mathtools, float}
\usepackage{lineno,hyperref,epstopdf,color}
\usepackage{tikz}
\usetikzlibrary{shapes,arrows}
\usepackage{fancyhdr}

\newcommand{\B}[1]{{\bm #1}}

\newcommand{\kv}{$k$-vector}
\newcommand{\ndkv}{$n$-dimensional $k$-vector}
\newcommand{\Ndkv}{$N$-dimensional $k$-vector}
\newcommand{\NDKV}{$N$-Dimensional $K$-Vector}

\fancypagestyle{firststyle}
{
	\fancyhf{}
	\fancyhead[RE,LO]{Applied Mathematics and Computation, Vol. 372, 2020. \\
		doi: 10.1016/j.amc.2019.125010}
	\fancyfoot[RE,LO]{doi: 10.1016/j.amc.2019.125010}
	\fancyfoot[LE,RO]{Page \thepage}
}

\begin{document}

\title{The \ndkv\ and its application to orthogonal range searching}

\author{David Arnas\thanks{Assistant Professor, Universidad de Zaragoza, Valentin Carderera 4, 22003 Huesca, Spain. Email: \textsc{darnas@unizar.es}}, \, Carl Leake\thanks{Ph.D. Graduate Student, Aerospace Engineering, Texas A\&M University, College Station, TX. E-mail: \textsc{leakec@tamu.edu}}, \, Daniele Mortari\thanks{Professor, Aerospace Engineering, Texas A\&M University, College Station, TX. E-mail: \textsc{mortari@tamu.edu}}}

\date{}	

\maketitle

\thispagestyle{firststyle}

\begin{abstract}
    This work focuses on the definition and study of the \ndkv, an algorithm devised to perform orthogonal range searching in static databases with multiple dimensions. The methodology first finds the order in which to search the dimensions, and then, performs the search using a modified projection method. In order to determine the dimension order, the algorithm uses the \kv, a range searching technique for one dimension that identifies the number of elements contained in the searching range. Then, using this information, the algorithm predicts and selects the best approach to deal with each dimension. The algorithm has a worst case complexity of $\mathcal{O}(nd(k/n)^{2/d})$, where $k$ is the number of elements retrieved, $n$ is the number of elements in the database, and $d$ is the number of dimensions of the database. This work includes a detailed description of the methodology as well as a study of the algorithm performance. 
\end{abstract}

\section{Introduction}

Orthogonal range searching is a very common problem that arises in a wide variety of applications including engineering, physics, mathematics, computational science, economics, and statistics. The problem consists of finding all the elements from a database that are contained in a given orthogonal range. In its simplest form, orthogonal range searching requires determining whether or not each element of the database is included in a given searching range. This can be done in general by checking all the elements from the database individually using a brute force approach. However, following this procedure, the time required by the algorithm quickly increases with the size of the database, making the algorithm slower as the size of the database increases. Thus, and in order to improve this limiting behavior, a wide variety of methodologies have been proposed to deal with the problem. Examples include the projection method~\cite{basic}, the grid files method~\cite{grid}, the $b$-tree~\cite{btree}, the quad tree~\cite{quadtree}, the $k$-d tree~\cite{Bentley}, the $R$-tree~\cite{rtrees}, and the $Kdb$-tree~\cite{kdbtree}; however, there are many others~\cite{willard,lueker,alstrup,arya}. The multitude of techniques shows that range searching is an extensively studied subject in computer science~\cite{agarwal,Chazelle}.

In this work, we introduce the \ndkv, a very efficient methodology for problems that require extensive orthogonal range searches in static databases with multiple dimensions. The \ndkv\ is based on the idea of generating an auxiliary database during preprocessing that contains information about the element distribution in each dimension. Afterwards, during the searching process, the algorithm uses this information to provide the number of elements in the searching range for all the dimensions individually, as well as the positions of these elements in the database. That way, and for each search, the algorithm finds the sequence of dimensions that is most likely to minimize the number of range comparisons, starting with the dimension with the smallest number of elements retrieved. Then, in each subsequent dimension, the algorithm continues with a brute force approach or an intersection approach depending on the situation. 

The \ndkv\ is the evolution, for databases in multiple dimensions, of the so called \kv, a range searching technique for one dimensional databases. The \kv\ range searching algorithm can be thought of as a hash table like algorithm based on a bijective hash function. It was originally developed for the identification of stars observed by wide field-of-view star trackers using computationally limited processors on board spacecraft~\cite{Original}. Since then, the \kv\ has been repeatedly and successfully validated in space~\cite{HETE,DraperPyramid} becoming part of the state-of-the-art algorithm, Pyramid~\cite{Pyramid}. In addition, the one-dimensional \kv\ has been successfully applied to solve other problems, including: inverting nonlinear functions~\cite{InvFun} (even Diophantine), generating random numbers with any prescribed distribution~\cite{David} (analytical or tabulated), solving for the intersection of nonlinear functions, identifying iso-surfaces for level-set analysis, and finding gene sequences in long DNA chains~\cite{Rogers}. 

The \kv\ is based on the idea of describing the nonlinearities of a sorted database using a vector of integers of a chosen size called the \kv~\cite{Adaptive} (a hash-like function). This is done by comparing the database distribution with a mapping function, for instance, a line. The key feature of the \kv\ is that the searching time complexity of the algorithm is independent of the database size; it only depends on the nonlinearity of the sorted database with respect to the mapping function~\cite{Neta}. As with the binary search technique, the one-dimensional $k$-vector only works with sorted databases. For this reason, a direct extension of this technique to $n$-dimensions is not possible due to the lack of clear ordering in multi-dimensional spaces. Therefore, a different approach is required, which is the focus of this work, the \ndkv. 

This manuscript is organized as follows. First, an overview of the algorithm is introduced, which aims to provide a general idea of the methodology and its process. Then, a detailed exposition of the methodology is presented, including the structure of the database and the auxiliary databases required (the index array, the \kv\ array, and the \kv\ line array). Afterwards, the performance study of the algorithm is assessed in terms of complexity and speed, and compared with brute force and $k$-d tree, which are other common orthogonal range searching algorithms. Next some modifications to the one-dimensional \kv\ algorithm are presented that show up as a result of the development of the \ndkv. Finally, some possible modifications to the algorithm are studied that reduce the memory required while maintaining the algorithm complexity.

\section{General overview of the algorithm}

The \ndkv\ is a numerical algorithm specifically devised to perform orthogonal range searches in multidimensional static databases. The algorithm first finds the most convenient dimension in which to perform a projection of the database by estimating the number of expected retrieved elements in each dimension. This is done using three auxiliary databases that first estimate the number of elements in the searching range in each dimension, and then, identify which points are contained in the projection. Since up to this point the elements identified have only been checked in one dimension, a brute force approach follows to assess the remaining dimensions. Therefore, only a fraction of the database is checked during the searching process. 

\tikzstyle{block} = [rectangle, draw, fill=white, text width=5.5em, text centered, rounded corners, minimum height=4em]
\tikzstyle{block2} = [rectangle, draw, fill=white, text width=7em, text centered, rounded corners, minimum height=4em]
\tikzstyle{line} = [draw, -latex']

\begin{figure}[!h]
	\centering
	\begin{tikzpicture}[node distance = 2.5cm]
	% Place nodes
	\node [block] (database) {Database};
	\node [block, right of=database] (Sorting1) {Sort in last dimension};
	\node [block, right of=Sorting1] (Subdatabase) {Subdatabase generation};
	\node [block, right of=Subdatabase] (eachsubdatabase) {For each subdatabase};
	\node [block, right of=eachsubdatabase] (Sorting2) {Sort in first dimension};
	% Draw edges
	\path [line] (database) -- (Sorting1);
	\path [line] (Sorting1) -- (Subdatabase);
	\path [line] (Subdatabase) -- (eachsubdatabase);
	\path [line] (eachsubdatabase) -- (Sorting2);
	\end{tikzpicture}
	\caption{Generation of the database structure}
	\label{flowchart:datastructure}
\end{figure}
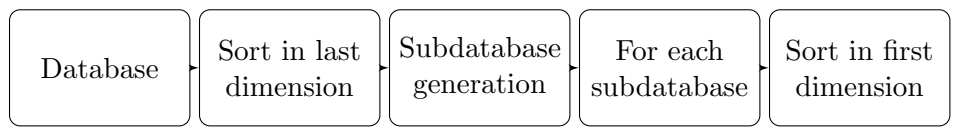

As such, the \ndkv\ can be regarded as a modified projection method in one dimension, where an auxiliary database is used rather than a binary search. However, an additional idea is introduced in the algorithm. Instead of searching in the whole database at the same time, the \ndkv\ performs multiple searches in subsets of the database, which are defined in such a way that they only contain elements in a known range from a given dimension selected during the preprocessing. This allows the \ndkv\ to asses two dimensions at the same time: the first one due to the structure of the database, and the second one due to the projection method.

The algorithm requires a one-time preprocessing effort that prepares the necessary auxiliary databases. During this preprocessing effort, the algorithm first sorts the whole database in a chosen dimension, for instance, the last dimension. Then, this sorted database is separated into sub-databases, where each sub-database contains all the elements that are within a given range in the selected dimension in such a way that each sub-database contains approximately the same number of elements. Afterwards, each sub-database is sorted again in a different dimension, for example, the first dimension, to generate the final data structure of the algorithm. A flowchart of this process can be seen in Figure~\ref{flowchart:datastructure}.

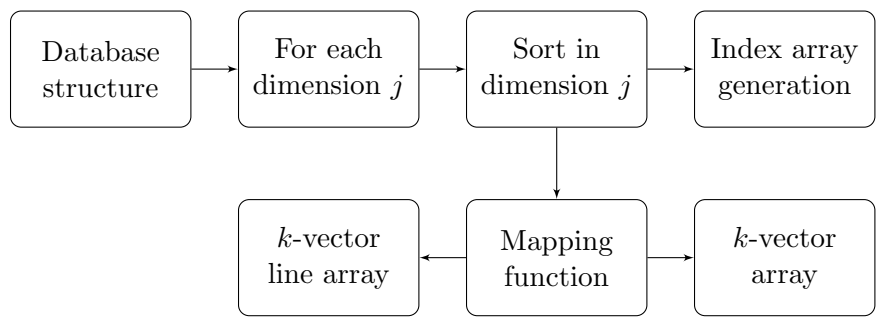
\begin{figure}[!h]
	\centering
	\begin{tikzpicture}[node distance = 3cm]
	% Place nodes
	\node [block] (database) {Database structure};
	\node [block, right of=database] (dimension) {For each dimension $j$};
	\node [block, right of=dimension] (sorting) {Sort in dimension $j$};
	\node [block, right of=sorting] (index) {Index array generation};
	\node [block, below of=sorting,  node distance=2.5cm] (map) {Mapping function};
	\node [block, right of=map] (kv) {\kv\ array};
	\node [block, left of=map] (line) {\kv\ line array};
	% Draw edges
	\path [line] (database) -- (dimension);
	\path [line] (dimension) -- (sorting);
	\path [line] (sorting) -- (index);
	\path [line] (sorting) -- (map);
	\path [line] (map) -- (line);
	\path [line] (map) -- (kv);
	\end{tikzpicture}
	\caption{Preprocessing process}
	\label{flowchart:kvector}
\end{figure}

After the data structure is built, three auxiliary databases are generated: 1) the index array, 2) the \kv\ array, and 3) the \kv\ line array. The index array contains information about the ordering of points in the different dimensions of the database. This index is used to map the database from a sorted order in a given dimension to the ordering of the data structure. 
The \kv\ array is used to quickly identify the database indexes that lay within a defined searching range in a given dimension. This is done by storing the indexes of the first elements, in a sorted database, whose components are above a prescribed set of known values in each dimension. These prescribed values are generated based on a linear mapping function, because a line can easily be inverted to obtain a one-to-one mapping. The elements required to invert this mapping function, the slope and the intercept, are stored in the \kv\ line array. Thus, the \kv\ line array is used to relate the searching range with the positions in the \kv\ array. A summary of this preprocessing effort is presented in Figure~\ref{flowchart:kvector}.

Once the preprocessing is finished, the \ndkv\ is able to perform the orthogonal search. This is done by following these steps, which are also presented schematically in Figure~\ref{flowchart:searching}:
\newpage
\begin{figure}[!h]
	\centering
	\begin{tikzpicture}[node distance = 4cm]
	% Place nodes
	\node [block2] (range) {Searching range};
	\node [block2, below of=range, node distance=2.5cm] (subdata) {Applied to each sub-database};
	\node [block2, right of=subdata] (dimension) {For each dimension $j$};
	\node [block2, right of=dimension] (map) {Mapping function};
	\node [block2, below of=map, node distance=2.5cm] (kv) {\kv};
	\node [block2, left of=kv] (number) {Number of elements in the projection};
	\node [block2, below of=subdata, node distance=2.5cm] (best) {Selection of dimension};
	\node [block2, below of=best, node distance=2.5cm] (projection) {Elements in the projection};
	\node [block2, below of=projection, node distance=2.5cm] (brute) {Brute force};
	\node [block2, right of=projection] (index) {Index array};
	% Draw edges
	\path [line] (range) -- (subdata);
	\path [line] (subdata) -- (dimension);
	\path [line] (subdata) -- (best);
	\path [line] (best) -- (projection);
	\path [line] (projection) -- (brute);
	\path [line] (dimension) -- (map);
	\path [line] (map) -- (kv);
	\path [line] (kv) -- (number);
	\path [line] (number) -- (best);
	\path [line] (kv) |- (index);
	\path [line] (index) -- (projection);
	\end{tikzpicture}
	\caption{Searching process}
	\label{flowchart:searching}
\end{figure}
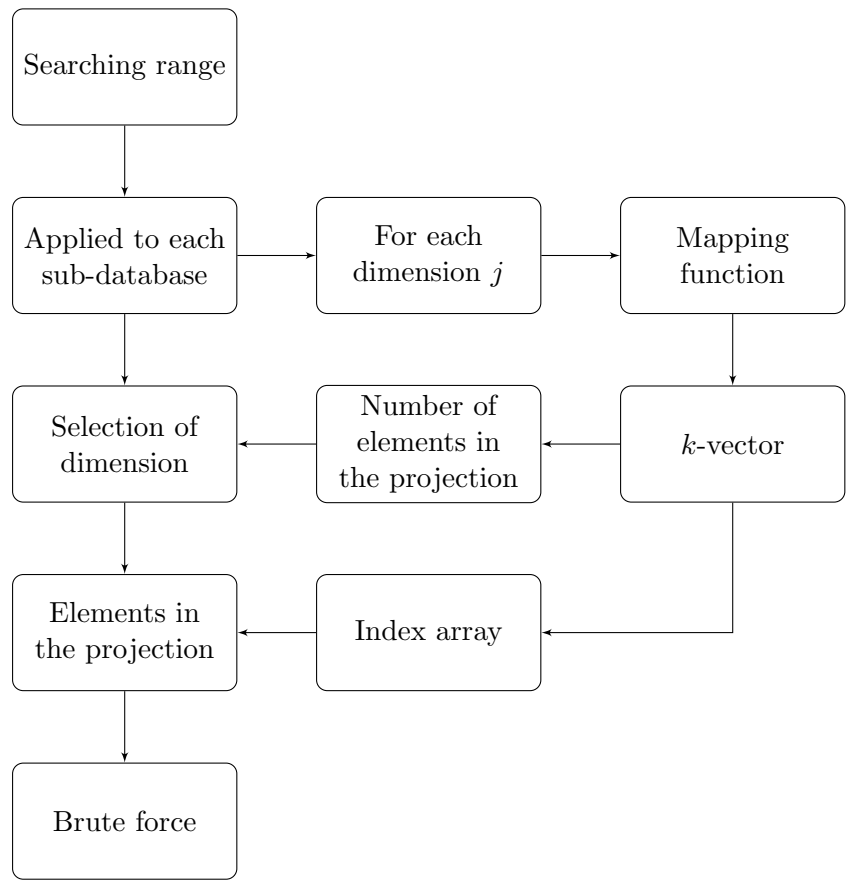
\begin{enumerate}
    \item \textbf{Select a sub-database.} The algorithm selects one of the sub-databases in which the database is distributed.
	\item \textbf{Apply mapping functions.} For each dimension of the problem, the algorithm applies the mapping functions to the user-specified searching ranges in order to obtain the approximate ranges in the \kv\ array.
	\item \textbf{Estimate the number of elements per dimension.} The algorithm accesses the \kv\ array and retrieves the range of indexes that are within the approximate ranges. Based on the design of the \kv\ array and one-dimensional \kv\ searching process, the approximated range is guaranteed to contain all of the elements in the user-specified range; however, it may also contain extraneous elements due to the range approximation performed. That way, the algorithm can now easily compute the approximate number of elements contained inside the range projection in each dimension.
	\item \textbf{Sort the dimensions.} All the dimensions are sorted based on the approximate number of points calculated during the previous step. This information is used to determine the dimension whose searching range contains the smallest number of elements. In addition, this information can be used later to define the order in which to evaluate the different dimensions of the problem during the brute force approach.
	\item \textbf{Remove extraneous elements.} A dimension is selected based on the previous information. Then, a local search is performed at the extreme elements of the approximated range to remove any extraneous elements.
	\item \textbf{Retrieve the elements.} Using the information provided by the \kv, the subset of elements from the sub-database inside the range in the selected dimension is retrieved using the index array.
	\item \textbf{Perform brute force.} A brute force approach is performed on the elements retrieved by the \kv\ that checks the dimensions that have not been evaluated yet. Using the information previously computed, the order in which the dimensions are checked is defined by the approximate number of points in each dimension.
	\item \textbf{Return to step 1.} Another sub-database is selected and the process is repeated until the whole database has been searched.
\end{enumerate}

In the following sections, we describe the data structure of the algorithm, the auxiliary databases, and the searching process in more detail. After presenting these sections, discussing the algorithm complexity, and comparing it with other common orthogonal range searching algorithms, we study the special case of one dimensional databases and discuss possible modifications of the algorithm that focus on reducing the memory required for the algorithm.

\section{Data structure}

The data structure selected for the \ndkv\ has two objectives. First, it positions the elements that belong to the same regions in the searching space into close or adjacent memory. Second, it provides a way to easily identify subsets of elements that are not contained in the searching range. This is done by performing two sorting processes in two different dimensions of the database. In that regard, and without loss of generality, we will assume that the first sorting process is performed in the last dimension of the database, while the second is performed in the first dimension of the database. This selection is only made to simplify the notation used.

The first step to generate the data structure is to sort all the elements of the database with respect to the last component in ascending order. Let $d$ be the number of dimensions of the problem. Then, the database is sorted in such a way that:
\begin{equation}
x_{i-1}(d-1) \leq x_i(d-1) \leq x_{i+1}(d-1) \quad \forall\ i\in\{1,\dots,n-2\}, 
\end{equation}
where $x_i(d-1)$ is the component in the last dimension of the point $x_i$, and $n$ is the number of elements in the database (ant thus, $i=0$ and $i=(n-1)$ are the first and last elements of the database). 

Once this sorting process is performed, this database is separated into sub-databases of nearly equal numbers of elements. Let $n_{db}$ be the number of sub-databases that are defined for a given database. Then, and since it is not always possible to distribute elements evenly in all sub-databases, we select the first sub-database to be the only one that has a different number of elements. That way, the number of points in a general sub-database is $n_p = \lfloor n/n_{db} \rfloor$, where $\lfloor x \rfloor$ is the largest integer less than the value of $x$ (i.e. the floor function), while $n_p' = n - n_p(n_{db}-1)$ is the number of elements in the first sub-database. By doing so, we are assuring that all the elements inside a given sub-database have their component $d$ ranging between the last components of the first and last elements of that sub-database, because the database was originally sorted in the last dimension. Mathematically, this is written as,
\begin{equation}
x_{p}(d-1) \leq x_i(d-1) \leq x_{q}(d-1) \quad \forall\ i\in\{p,\dots,q\},
\end{equation}
where $p$ and $q$ are the first and last elements of a given sub-database respectively. At this point, it is important to note that the choice of the number of sub-databases, $n_{db}$, has a large impact on the performance of the algorithm. This is due to the fact that $n_{db}$ is directly related to the number of reference values generated for the last dimension (which control the precision of the range approximation during the searching process), but also because it determines the number of different searches that will be performed later (one for each sub-database). This means that $n_{db}$ can be used to adapt the algorithm to the particularities of the problem considered.

The objective now is to provide a sorted structure in a different dimension. As defined before, each sub-database contains all the points inside a known range in the last dimension. Therefore, if another sorting process is performed in a different dimension, the elements from the same hyperprism will be located in adjacent positions in memory. In order to do that, this second sorting is performed in each independent sub-database, but this time with respect to the first dimension. This means that the elements within a range in the first dimension are adjacent in memory for each sub-database. This property is used by the algorithm to improve its search performance when possible.

With that, the database structure of the algorithm is set. In the following section, the auxiliary databases are presented, which improve the search performance of the algorithm. However, these auxiliary databases are in fact optional, as it is possible to perform the range searching process with just the data structure as presented in Section~\ref{sec:options}.

\section{Auxiliary databases}

As part of the preprocessing effort of the \ndkv, the algorithm generates a set of auxiliary databases whose main objective is to provide information on the database distribution. Later, during the searching process, the algorithm will use this information to quickly locate the elements from the database that lay inside a given searching range in one of the database dimensions.

The \ndkv\ makes use of three auxiliary databases: the index array, the \kv\ array, and the \kv\ line array. These auxiliary databases are computed independently for each sub-database defined in the data structure. This means that during the searching process, each sub-database can be searched independently without further information. This property is interesting, as it may be useful for parallelizing the \ndkv\ algorithm. In the following subsections, these auxiliary databases are studied in more detail.

\subsection{Index array}

The index array is an auxiliary database that contains information on how the points are sorted with respect to the different dimensions of the problem. To generate it, a sorting process in ascending order must be done for each dimension. Let $j\in\{0,\dots,d-1\}$ be a given dimension of the problem, and let $\B{y_j}$ be the array containing the components of all the points in that dimension. Then, we define $\B{s_j}$ to be the components of $\B{y_j}$ sorted in ascending order. This means that:
\begin{eqnarray}\label{eq:sinc}
& \B{s_j}(i) \leq \B{s_j}(i+1), & \quad \forall\ i\in\{0,\dots,n-2\}; \nonumber \\
& \B{s_j}(i) = \B{y_j}(\B{I_j}(i)), & \quad \forall\ i\in\{0,\dots,n-1\};
\end{eqnarray}
where $\B{I_j}$ is the index array. The index array relates the positions of the elements from the original database $\B{y_j}$ with respect to the sorted one $\B{s_j}$ in the dimension $j$. The sorting and computation of $\B{I_j}$ has a complexity of $\mathcal{O}(n \log n$) and represents the most demanding process during the preprocessing of the \ndkv. Moreover, it is important to note that the index array for any dimension is defined with respect to the same data structure, and thus, a common reference is shared for all the dimensions of the problem.

The sorting process is repeated for all the dimensions of the problem, storing the values of $\B{I_j}$ to generate the complete index array $\B{I}$. Once the preprocessing is finished, the values $\B{s_j}$ can be erased to free up memory, as they are not used during the searching process. Note that since the number of points and the number of dimensions is the same as the original database, but the first dimension is already sorted in each sub-database, the size of the index array is $n(d-1)$ integer numbers.

\subsection{\kv\ array}

The \kv\ array is an auxiliary database that contains information on how the element components in each dimension are distributed. This auxiliary database requires defining a mapping function that will serve as a reference to which the database distribution will be compared. This mapping function performs a one-to-one application (a bijection) between the index values (the location of elements in the database) and a set of prescribed values in the searching space defined by the mapping function. In other words, the mapping function serves as a first approximation between the sorted values of the database elements and their relative location, that is, their index. In addition, the mapping function is selected in such a way that it is easy to invert to improve search performance. 

In order to generate this auxiliary database, we first need to decide the size that this database will have. This is an important parameter not only because it will determine the size of this auxiliary database and the number of operations required to compute it, but also because it affects the search performance of the algorithm. In particular, having a large size implies that the retrieval of elements will be faster since the algorithm will have more resolution in the grid defined in each dimension. Conversely, having a lower size implies that the probability of obtaining extraneous elements from the range increases. This means that the size of the \kv\ array can be used as a design parameter to adapt this methodology to fit the needs of the problem to be solved.

Let $n_k$ be the size of the \kv\ array for a given dimension $j\in\{0,\dots,d-1\}$, and let $z_j(i)$ with $i\in\{0,\dots,n_k-1\}$ be the mapping function defined for that dimension. Then, the \kv\ array for that dimension, $\B{k_j}$, can be described as:
\begin{equation}\label{eq:kvec1}
\B{k_j}(i) = r \quad \Big| \quad \B{s_j}(r)<\B{z_j}(i)\leq\B{s_j}(r+1),
\end{equation}
where $r\in\{0,\dots,n_p-2\}$ and $n_p$ is the number of elements of the sub-database. This means that $\B{k_j}(i)$ stores the number of elements in the database smaller than the corresponding value of the mapping function at that same $i$-th index location, $\B{z_j}(i)$, or in other words, it contains the index of the first element that fulfills the relation presented in Eq.~\eqref{eq:kvec1}. This computation has to be done for all the dimensions of the problem, storing the information contained in $\B{k_j}$ in the \kv\ array, $\B{k}$. As it can be seen, the generation of the \kv\ array only requires reading the database and comparing each element component with the auxiliary function. This leads to a complexity of $\mathcal{O}(nd)$ to generate the whole \kv\ array, and a memory requirement of $n_kd$ integer numbers for each sub-database.

\begin{figure}[!h]
	\centering
	\includegraphics[width=0.80\linewidth]{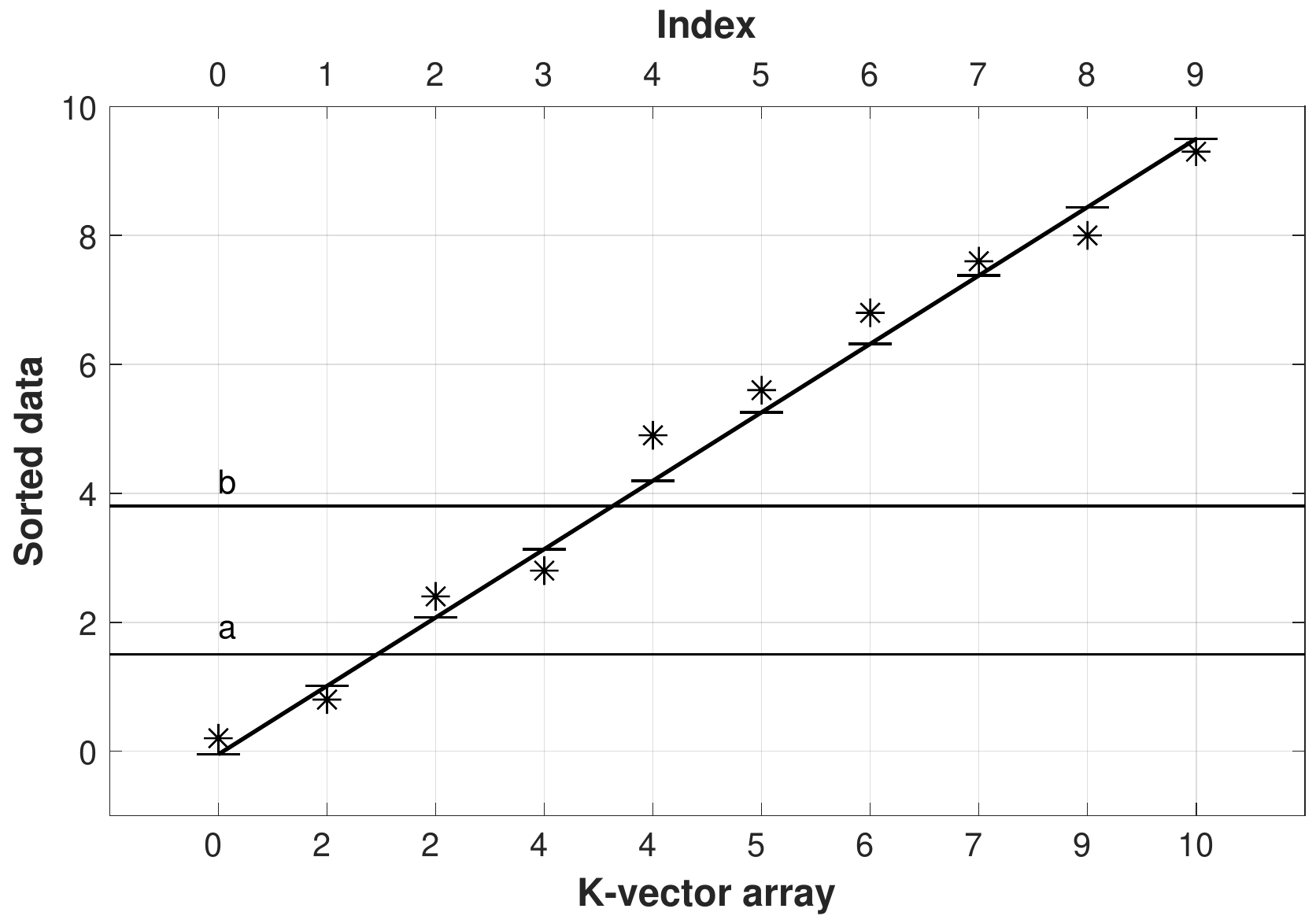}
	\caption{One dimensional \kv\ example}
	\label{fig:1dkv}
\end{figure}

Figure \ref{fig:1dkv} graphically shows how the \kv\ array, index, and sorted database are connected. In Fig. \ref{fig:1dkv}, the sorted database points are shown as asterisks, the discrete mapping function values are shown as short, horizontal lines, and the line that connects the discrete mapping function values is the mapping function. In addition, an example search query from $[a,b]$ is shown via two long horizontal lines. The $x$-axis labels on top of the figure show the index value for each of the sorted data points, and the $x$-axis labels on the bottom of the figure show the \kv\ array values for each of the discrete mapping function points. The $y$-axis gives the value of each of the sorted data points, and the value of the each of the mapping function points.

\subsection{\kv\ line array}

The \kv\ line array is an auxiliary database that directly relates a given searching range and its associated boundary elements in the \kv\ array. This means that both auxiliary databases (the \kv\ array and the \kv\ line array) always work in cooperation. 

In order to generate the \kv\ line array, the mapping function of the \kv\ must first be defined. In general, the mapping function is selected as a line function due to its easy and fast inversion. In this case, the \kv\ provides general information about the non-linearity of the database. The definition of this mapping line is performed by taking the value of the mapping function as the variable of the function, that is: 
\begin{equation}
i = m_j\B{z_j}(i)+q_j, \quad \forall \ i\in\{0,\dots,n_k-1\}. 
\end{equation}
This is done in order to remove the requirement of inverting the mapping function during the searching process, thereby, improving the speed of the methodology. The mapping function has the following expression:
\begin{equation}\label{eq:z}
\B{z_j}(i) = \displaystyle\frac{i}{m_j} - \frac{q_j}{m_j}, \quad \forall \ i\in\{0,\dots,n_k-1\}, 
\end{equation}
which represents a line connecting the extreme elements of the sorted database for the dimension $j$. However, and in order to take into account possible rounding and machine errors, the range of this line is slightly extended to the point-pairs $(1,\ \B{s_j}(l) - \delta\varepsilon)$ and $(n_k,\ \B{s_j}(u) + \delta\varepsilon)$, where
\begin{equation}
\delta\varepsilon = (n_k - 1) \varepsilon,
\end{equation}
$\varepsilon$ is the relative machine precision, and $\B{s_j}(l)$ and $\B{s_j}(u)$ are the minimum and maximum values in the sub-database for each dimension. Therefore, $m_j$ and $q_j$ can be defined as:
\begin{eqnarray}\label{eq:mandq}
m_j & = & \dfrac{n_k - 1}{\B{s_j}(u) - \B{s_j}(l) + 2\delta\varepsilon}, \nonumber \\
q_j & = & - \dfrac{n_k - 1}{\B{s_j}(u) - \B{s_j}(l) + 2\delta\varepsilon}(\B{s_j}(l) - \delta\varepsilon),
\end{eqnarray}
which are the parameters that are stored in the \kv\ line array. This means that, using a line as a mapping function, $2dn_{db}$ real numbers are required to be stored in this auxiliary database for each sub-database defined. Moreover, the algorithm complexity required to generate the complete \kv\ line array is $\mathcal{O}(n_{db}d)$.

\section{Searching process}

The first step in the searching process of the \ndkv\ is to determine the boundary indexes of the \kv\ line array for the different ranges in each dimension. This is done using the mapping functions. Let $j\in\{0,\dots,d-1\}$ be a given dimension in which the range $[a_j,b_j]$ is defined. Then, the approximate boundary indexes in the \kv\ line array are:
\begin{eqnarray} \label{eq:ajbj}
A_j & = & \lfloor m_ja_j + q_j \rfloor, \nonumber \\
B_j & = & \lceil m_jb_j + q_j \rceil,
\end{eqnarray}
where $A_j$ and $B_j$ the lower and upper boundaries in the \kv\ array for the dimension $j$, and $\lceil x \rceil$ is smallest integer larger than $x$ (i.e. the ceil function). A visual example of this process is shown in Fig. \ref{fig:1dkv}.

Instead of performing the original search defined by $[a_j,b_j]$, the algorithm is actually searching for the best approximation of the range using the prescribed values generated by the mapping function. Note that these approximated ranges are selected in such a way that the defined range $[a_j,b_j]$ is always contained in the approximation. Later, a small search will be performed on the extreme elements of the range, but for the moment, we are only interested in having an approximated value of the number of elements contained in the range of each dimension. In that regard, the number of points within this approximated range in the dimension $j$, $\B{p}(j)$, can be obtained using the following expression:
\begin{equation}\label{eq:pj}
\B{p}(j) = \B{k_j(B_j)} - \B{k_j(A_j)}.
\end{equation}
Then, if the operations defined in Eqs.~\eqref{eq:ajbj} and~\eqref{eq:pj} are repeated for all the dimensions of the database, it is possible to sort the dimensions based on the number of points expected to be retrieved. This is done by applying a sorting algorithm to array $\B{p}$ (of size $d$), to obtain a sorted array $\B{g}$ in ascending order.

It is important to note that since the number of points in the searching range of each dimension is evaluated, there will be entire sub-databases where no retrieval will be performed. Particularly, these sub-databases correspond to the ones whose elements are outside the searching range of the last dimension, and thus, they can be removed from the searching range without any problem. This also means that the algorithm does not need to continue the searching process in these sub-databases. As a result, this property, derived from the data structure, allows performing an active search in two dimensions instead of just the one (which is provided by the \kv\ methodology).

At this point, we already know which dimension has the smallest number of elements in its searching range. In general, this should be the dimension in which the algorithm should continue the searching process by using the information provided by the \kv\ array. However, this is not always true. The reason for this is that checking the searching range is faster when the elements studied are adjacent in memory. Therefore, there could be cases where continuing the searching process in the first dimension of the problem is faster despite having to evaluate a larger number of elements compared with other dimensions. For instance, with the test computers used by the authors, the ratio in speed between reading the database in an ordered process and accessing the database in a random order is on average:
\begin{equation}\label{eq:gain}
r = \displaystyle\frac{3}{2}(log_{10}(n_p) - 3),
\end{equation}
where $n_p \geq 10^3$ is the number of points to be studied (or in the case of the \ndkv, the number of elements in each sub-database). For this reason, the \ndkv\ decides which is the most convenient dimension to continue the searching process based on the relation between $\B{g}(0)$ and $\B{p}(0)$ (the first components of the sorted and unsorted arrays for approximate number of elements per dimension). That way, two options are available. If $\B{p}(0) > r\B{g}(0)$, the \ndkv\ continues the searching process in the dimension with the smallest number of elements. Conversely, the algorithm continues the searching process in the first dimension of the data structure. Regardless of the dimension that is selected in this step, the process followed afterwards is completely equivalent.

Once the most convenient dimension is selected, the algorithm first checks the real range of the dimension selected, and then, it continues with the rest of dimensions of the problem. In the following lines we study each case independently due to their differences.

Let $j$ be the dimension selected by the algorithm. Then, from the application of the \ndkv, we know that all the elements in the searching range are contained in the subset of points defined by these indexes:
\begin{eqnarray}
i \ | \ (A_j \leqslant i < B_j) & \quad & \text{if}\ j = 0, \nonumber \\
I_j(i) \ | \ (A_j \leqslant i < B_j) & \quad & \text{if}\ j \neq 0.
\end{eqnarray}
Therefore, we first have to remove the elements that are located outside the range $[a_j,b_j]$. In order to do that, and instead of checking all the elements individually, a linear search is performed on the extreme elements of the range by increasing the value of $A_j$ and decreasing the value of $B_j$ until the extreme elements are within the searching range. This means that for linear databases, on average, there will be $n_p/n_k$ elements that must be removed. If the database is highly non-linear, or the size of the \kv\ array is not large enough, this linear search can be substituted by a binary search. This binary search is performed between the elements $\B{k_j(A_j)}$ and $\B{k_j(A_j}+1)$ for the extreme elements at the lower end of the searching range, and between elements $\B{k_j(B_j}-1)$ and $\B{k_j(B_j)}$ for the extreme elements at the upper end of the searching range. Note also that the selection of the methodology to follow can be done during the searching process since we know number of elements in the extremes (given by $[\B{k_j(A_j}+1)-\B{k_j(A_j)}]$, and $[\B{k_j(B_j)}-\B{k_j(B_j}-1)]$ respectively), and the expected number of elements for a linear distribution ($n_p/n_k$). Therefore, if the number of elements in the extremes is several times larger than the expected, the algorithm should change to a binary search in order to remove the possible extraneous elements and improve its performance. 

The remaining dimensions are evaluated using a brute force approach, where the components in each dimension are checked following the ordering of dimensions defined by $\B{p}$ and $\B{g}$. This is done to slightly improve the search performance of the algorithm since it reduces the average number of range comparisons in the elements retrieved.

\section{Simple example of application}

This section provides an example of the data structure, auxiliary databases, and searching process for a simple database consisting of ten elements and three dimensions. The original database considered is shown in table \ref{tab:OrigData}, where the variables $x$, $y$, and $z$ are used to denote the first, second, and third dimension respectively. The objective is to identify the elements within the range $[2, 8]\times[5, 6]\times[1, 3]$, that is, $2\leq x\leq 8$, $5\leq y\leq 6$, and $1\leq z\leq 3$. 

\begin{table}[!h]
	\centering
	\caption{Original database}
	\label{tab:OrigData}
	\begin{tabular}{|c||c|c|c|c|c|c|c|c|c|c|}
		\hline
		$x$ & 6 & 9 & 0 & 2 & 4 & 3 & 5 & 1 & 8 & 7\\
		\hline
		$y$ & 9 & 3 & 2 & 7 & 1 & 0 & 6 & 8 & 4 & 5\\
		\hline
		$z$ & 1 & 9 & 5 & 3 & 4 & 0 & 2 & 8 & 6 & 7\\
		\hline
	\end{tabular}
\end{table}

\subsection{Preprocessing - Data structure}

First, the original database is sorted in the last dimension, as is shown in table \ref{tab:DataSort3}. Then, the original database sorted in the last dimension from table \ref{tab:DataSort3} is divided into sub-databases (two sub-databases in this case) and re-sorted in the first dimension. These sub-databases are shown in Table \ref{tab:subData}. Table \ref{tab:subData} also includes an element index that can be used to identify each element.

\begin{table}[!h]
	\centering
	\caption{Original database sorted in last dimension}
	\label{tab:DataSort3}
	\begin{tabular}{|c||c|c|c|c|c|c|c|c|c|c|}
		\hline
		$x$ & 3 & 6 & 5 & 2 & 4 & 0 & 8 & 7 & 1 & 9\\
		\hline
		$y$ & 0 & 9 & 6 & 7 & 1 & 2 & 4 & 5 & 8 & 3\\
		\hline
		$z$ & 0 & 1 & 2 & 3 & 4 & 5 & 6 & 7 & 8 & 9\\
		\hline
	\end{tabular}
\end{table}

\begin{table}[!h]
	\centering
	\caption{Sub-databases}
	\label{tab:subData}
	\vspace{0.05cm}
	\begin{minipage}{0.45\linewidth}
		\begin{tabular}{|c||c|c|c|c|c|}
			\hline
			index & 0 & 1 & 2 & 3 & 4 \\
			\hline
			\hline
			$x$ & 2 & 3 & 4 & 5 & 6\\
			\hline
			$y$ & 7 & 0 & 1 & 6 & 9\\
			\hline
			$z$ & 3 & 0 & 4 & 2 & 1\\
			\hline
		\end{tabular}
	\end{minipage}
	\begin{minipage}{0.45\linewidth}
		\begin{tabular}{|c||c|c|c|c|c|}
			\hline
			index & 5 & 6 & 7 & 8 & 9 \\
			\hline
			\hline
			$x$ & 0 & 1 & 7 & 8 & 9\\
			\hline
			$y$ & 2 & 8 & 5 & 4 & 3\\
			\hline
			$z$ & 5 & 8 & 7 & 6 & 9\\
			\hline
		\end{tabular}
	\end{minipage}
\end{table}

\subsection{Preprocessing - Auxiliary databases}

After creating the sub-databases shown in table \ref{tab:subData}, the auxiliary databases can be generated. The first auxiliary database is the index array. This data structure is used to sort each dimension of each sub-database in ascending order. Note that although the index array need not be generated for the first dimension because the sub-databases are sorted in the first dimension, it is included anyway for completeness. The index array is shown in table \ref{tab:indexArray}.

\begin{table}[!h]
	\centering
	\caption{Index array}
	\label{tab:indexArray}
	\vspace{0.05cm}
	\begin{minipage}{0.45\linewidth}
		\begin{center}
			\begin{tabular}{|c||c|c|c|c|c|}
				\hline
				$x$ & 0 & 1 & 2 & 3 & 4\\
				\hline
				$y$ & 1 & 2 & 3 & 0 & 4\\
				\hline
				$z$ & 1 & 4 & 3 & 0 & 2\\
				\hline
			\end{tabular}
		\end{center}
	\end{minipage}
	\begin{minipage}{0.45\linewidth}
		\begin{center}
			\begin{tabular}{|c||c|c|c|c|c|}
				\hline
				$x$ & 5 & 6 & 7 & 8 & 9\\
				\hline
				$y$ & 5 & 9 & 8 & 7 & 6\\
				\hline
				$z$ & 5 & 8 & 7 & 6 & 9\\
				\hline
			\end{tabular}
		\end{center}
	\end{minipage}
\end{table}

Then, the \kv\ line array can be generated according to Eq. \eqref{eq:mandq}, and the \kv\ array can be generated according to Eq. \eqref{eq:kvec1}. The \kv\ line array, rounded to two decimal places, is shown in table \ref{tab:kvLineArray}, and the \kv\ array is shown in table \ref{tab:kvArray}.

\begin{table}[!h]
	\centering
	\caption{$K$-vector line array}
	\label{tab:kvLineArray}
	\vspace{0.05cm}
	\begin{minipage}{0.45\linewidth}
		\begin{center}
			\begin{tabular}{|c||c|c|}
				\hline
				& $m$ & $q$ \\
				\hline
				\hline
				$x$ & 1.00 & -2.00\\
				\hline
				$y$ & 0.44 & 0.00\\
				\hline
				$z$ & 1.00 & 0.00\\
				\hline
			\end{tabular}
		\end{center}
	\end{minipage}
	\begin{minipage}{0.45\linewidth}
		\begin{center}
			\begin{tabular}{|c||c|c|}
				\hline
				& $m$ & $q$ \\
				\hline
				\hline
				$x$ & 0.44 & 0.00\\
				\hline
				$y$ & 0.67 & -1.33\\
				\hline
				$z$ & 1.00 & -5.00\\
				\hline
			\end{tabular}
		\end{center}
	\end{minipage}
\end{table}

\begin{table}[!h]
	\centering
	\caption{$K$-vector array}
	\label{tab:kvArray}
	\vspace{0.05cm}
	\begin{minipage}{0.45\linewidth}
		\begin{center}
			\begin{tabular}{|c||c|c|c|c|c|}
				\hline
				$x$ & 0 & 1 & 2 & 4 & 5\\
				\hline
				$y$ & 0 & 2 & 2 & 3 & 5\\
				\hline
				$z$ & 0 & 1 & 2 & 4 & 5\\
				\hline
			\end{tabular}
		\end{center}
	\end{minipage}
	\begin{minipage}{0.45\linewidth}
		\begin{center}
			\begin{tabular}{|c||c|c|c|c|c|}
				\hline
				$x$ &  5 &  7 &  7 &  7 & 10\\
				\hline
				$y$ &  5 &  7 &  8 &  9 & 10\\
				\hline
				$z$ &  5 &  6 &  7 &  9 & 10\\
				\hline
			\end{tabular}
		\end{center}
	\end{minipage}
\end{table}

\subsection{Searching process}

The first step is to find the approximate boundaries of the searching range. By using Eq.~\eqref{eq:ajbj} in the first sub-database:
\begin{eqnarray}
	A_x & = & \text{floor}\left( 1.00\cdot 2 - 2.00 \right) = 0 \longrightarrow \B{k_x(A_x)} = 0; \nonumber \\
	B_x & = & \ \text{ceil}\left( 1.00\cdot 8 - 2.00 \right) = 6 \longrightarrow \B{k_x(B_x)} = 5; \nonumber \\
	A_y & = & \text{floor}\left( 0.44\cdot 5 + 0.00 \right) = 2 \longrightarrow \B{k_y(A_y)} = 2; \nonumber \\
	B_y & = & \ \text{ceil}\left( 0.44\cdot 6 + 0.00 \right) = 3 \longrightarrow \B{k_y(B_y)} = 3; \nonumber \\
	A_z & = & \text{floor}\left( 1.00\cdot 1 + 0.00 \right) = 1 \longrightarrow \B{k_z(A_z)} = 1; \nonumber \\
	B_z & = & \ \text{ceil}\left( 1.00\cdot 3 + 0.00 \right) = 3 \longrightarrow \B{k_z(B_z)} = 4;
\end{eqnarray}
which using Eq.~\eqref{eq:pj} means that in dimension $x$ there are $5$ elements; in dimension $y$, $1$; and in dimension $z$, $3$ elements. Therefore, $y$ is the dimension selected to perform the projection. In this projection there is only one possible candidate, the point corresponding to the position $2$ in the sorted database with respect to dimension $y$, that is, the point with index $I_y(2)=3$ and coordinates $(5, 6, 2)$. Now, this point must be checked in the other dimensions, starting with dimension $z$ (the dimension with the lowest number of elements excluding dimension $y$), and continuing with dimension $x$. That way, we conclude that the point with index $3$ lies within the searching range.

Additionally, the algorithm must also check the second sub-database. By using Eq.~\eqref{eq:ajbj} again in this sub-database, we find that there are no elements in the projection of the searching range in dimension $z$ (since $\B{k_z(A_z)}=\B{k_z(B_z)}=5$). Thus, this sub-database does not contain any elements within the searching range. Therefore, the algorithm has finished the searching process.

\section{Algorithm complexity}

In order to estimate the algorithm complexity, we will assume a worst case scenario for the algorithm. Since the algorithm evaluates the number of elements in the range of each dimension, the worst case scenario for the \ndkv\ happens when the projection of the searching range in every dimension contains the same number of elements. In that regard, it is important to remember that the algorithm performs a projection of the database in two selected dimensions, the last one through the database structure, and another selected by the \kv. This means that, in general, the number of points that the algorithm must study is larger than the number of points in the searching range. Therefore, we need to evaluate this number of elements since it constitutes the largest computational effort made by the algorithm. 

Let $n$ be the number of elements of the database, $d$ the number of dimensions of the database, and $k$ the number of elements contained in the searching range. Then, and under the worst case scenario considered, the range in each dimension contains $n(k/n)^{(1/d)}$ elements, which are the ones that should be assessed if only a \kv\ is applied. Note that this value represents a fraction of the number of elements considered by the \kv\ based on the number of dimensions and the number of elements that have to be retrieved. However, the \ndkv\ also arranges the database into $n_{db}$ sub-databases based on the distribution in the last dimension, which allows the algorithm to remove the elements that are outside the searching range in the last dimension. This means that before applying the \kv\ the size of the database is reduced to $n(k/n)^{(1/d)}$ elements, and thus, the \kv\ only deals with this number of points instead of the whole database. Therefore, the number of elements that are evaluated by the algorithm is $n(k/n)^{(2/d)}$. Hence, due to the fact that the number of dimensions is $d$, the algorithm complexity is $\mathcal{O}(nd(k/n)^{(2/d)})$.

The previous study only considers the number of elements that are evaluated and not the number of operations required to obtain those elements. Therefore, we also study these computations in more detail and show that they represent a much smaller number of operations during the searching process. First, the algorithm requires transforming the searching range into the boundaries of the \kv\ line array. This is done in $\mathcal{O}(d)$ operations. Once this is completed, the dimensions are sorted based on the expected number of points retrieved. Using an algorithm such as quicksort of mergesort implies a complexity of $\mathcal{O}(d\log d)$. Finally, the algorithm must check if extraneous elements are contained in the data retrieved by the \kv\ for the chosen dimension. If a linear search is performed in the extreme elements, this will require an average of $n/(n_{db}n_k)$ operations. Note that if the size of the \kv\ array is equal to the size of the database, $n=n_{db}n_k$, there is only one extraneous element on average. Conversely, if a binary search is performed, the algorithm will take an average of $\log(n/(n_{db}n_k))$ operations. As it can be seen, all these processes are orders of magnitude smaller than the algorithm complexity derived before, and thus, they are considered negligible in the calculation of the algorithm complexity.

\section{Performance of the \NDKV}

The search times presented in this section were gathered in C using the Query Performance Counter function on a Windows 10 operating system using an Intel(R) Core(TM) i7-7700 CPU running at 3.60GHz with 16 GB of RAM. Moreover, in order to be conservative in the evaluation of the search performance of the \ndkv, the worst case scenario for this method has been considered; that is, there are approximately the same number of elements in the range of each dimension. This implies that the \ndkv\ cannot benefit from evaluating the dimensions in a particular order. In addition, the number of elements used in the \kv\ array for these tests was set to be a tenth the size of the number of elements in the database ($n_k=n/10$). 

\begin{figure}[!h]
    \centering
    \includegraphics[width=0.80\linewidth]{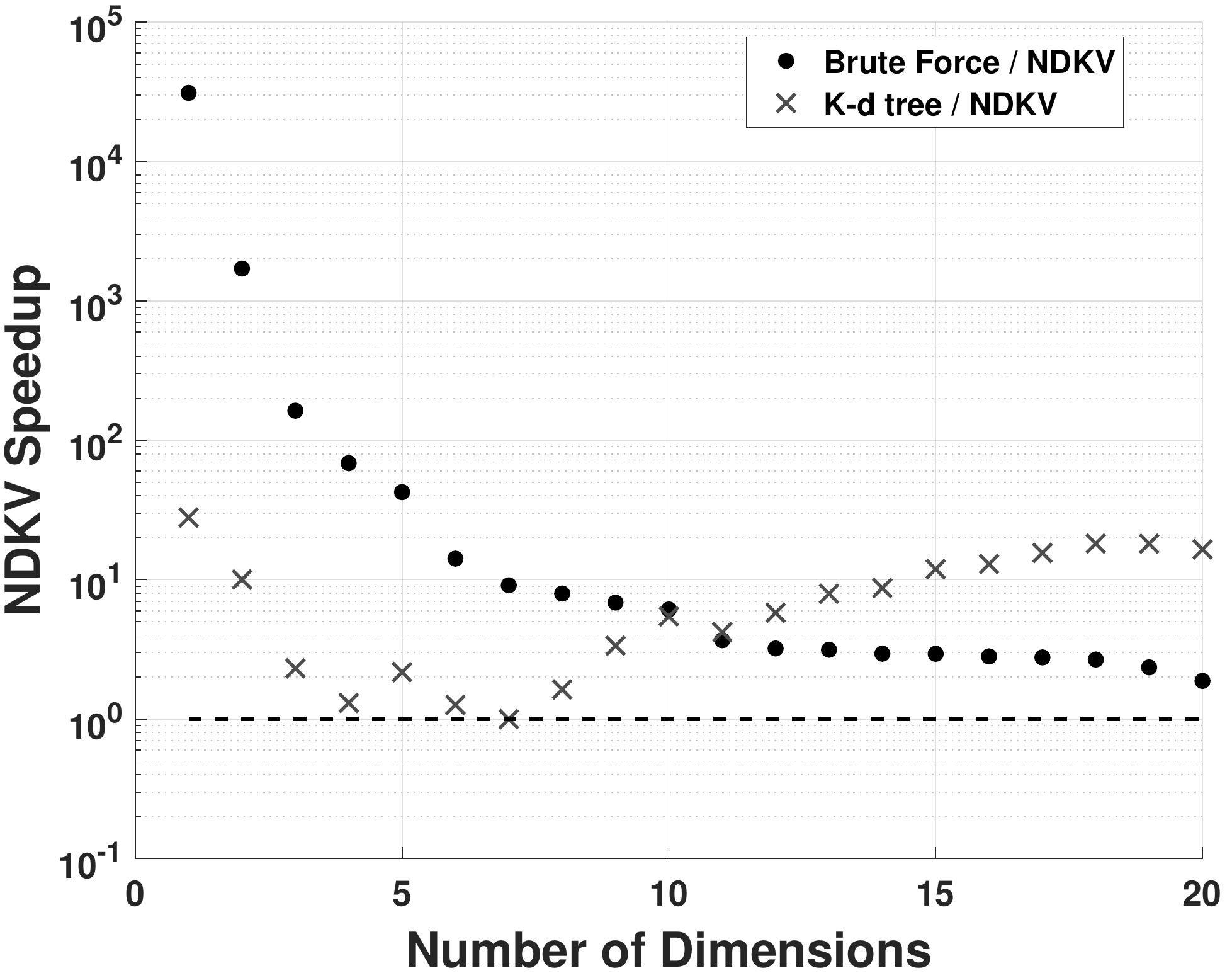}
    \caption{\Ndkv\ speedup versus number of dimensions for a database with one million elements and approximately 0.01\% of the database being retrieved}
    \label{fig:NumDimComp0.01}
\end{figure}

\begin{figure}[!h]
    \centering
    \includegraphics[width=0.80\linewidth]{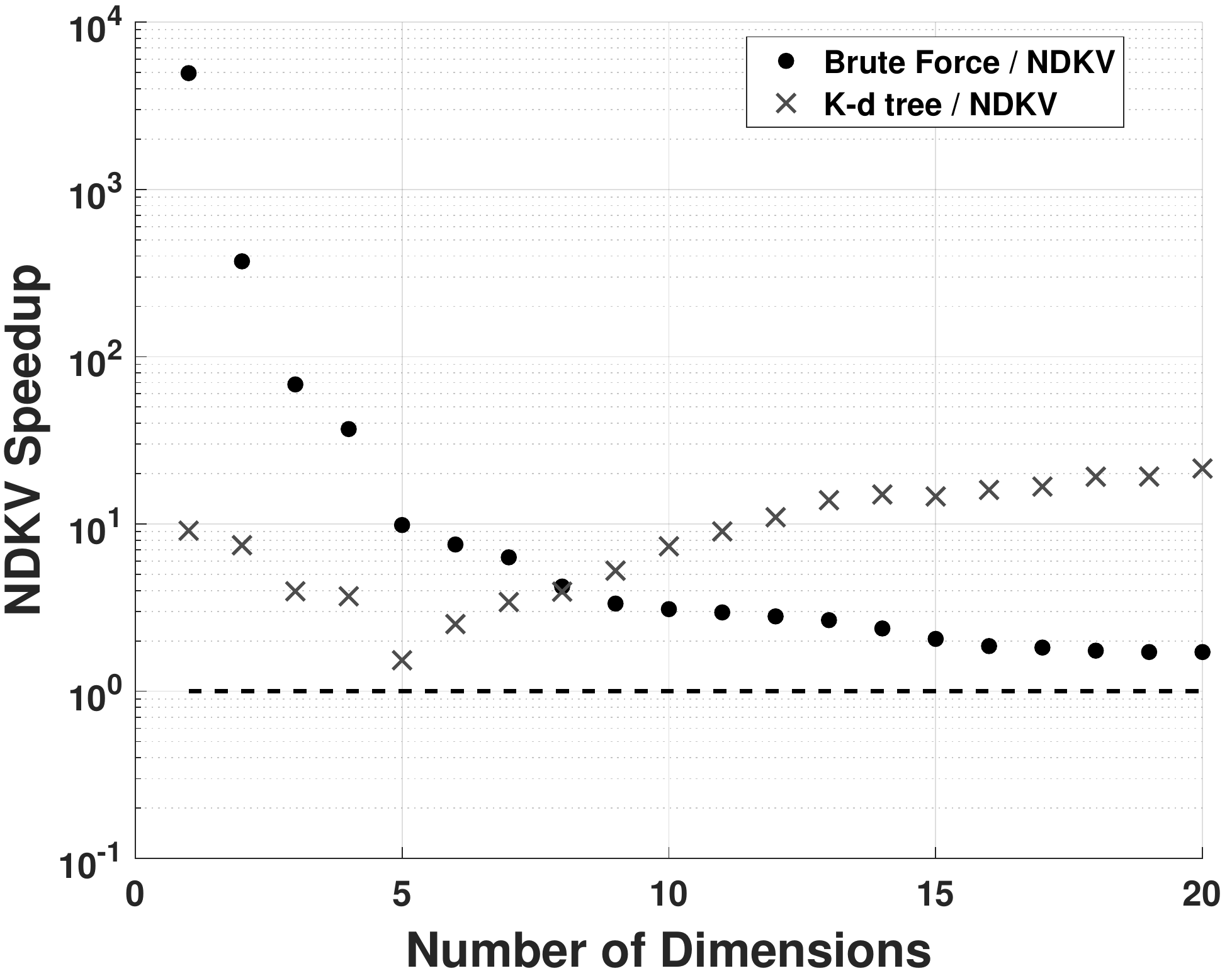}
    \caption{\Ndkv\ speedup versus number of dimensions for a database with one million elements and approximately 0.1\% of the database being retrieved}
    \label{fig:NumDimComp0.1}
\end{figure}

\begin{figure}[!h]
    \centering
    \includegraphics[width=0.80\linewidth]{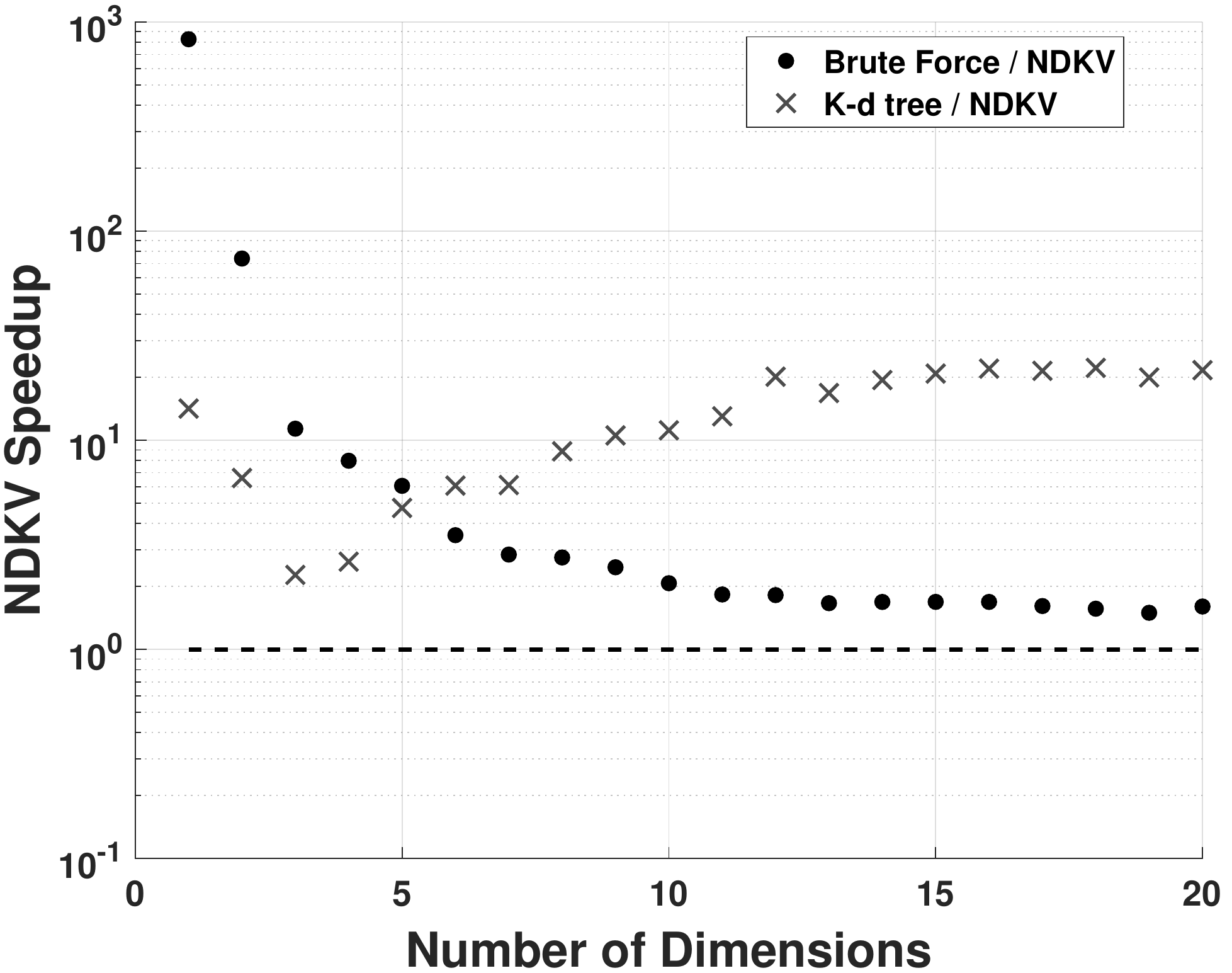}
    \caption{\Ndkv\ speedup versus number of dimensions for a database with one million elements and approximately 1\% of the database being retrieved}
    \label{fig:NumDimComp1.0}
\end{figure}

\begin{figure}[!h]
    \centering
    \includegraphics[width=0.80\linewidth]{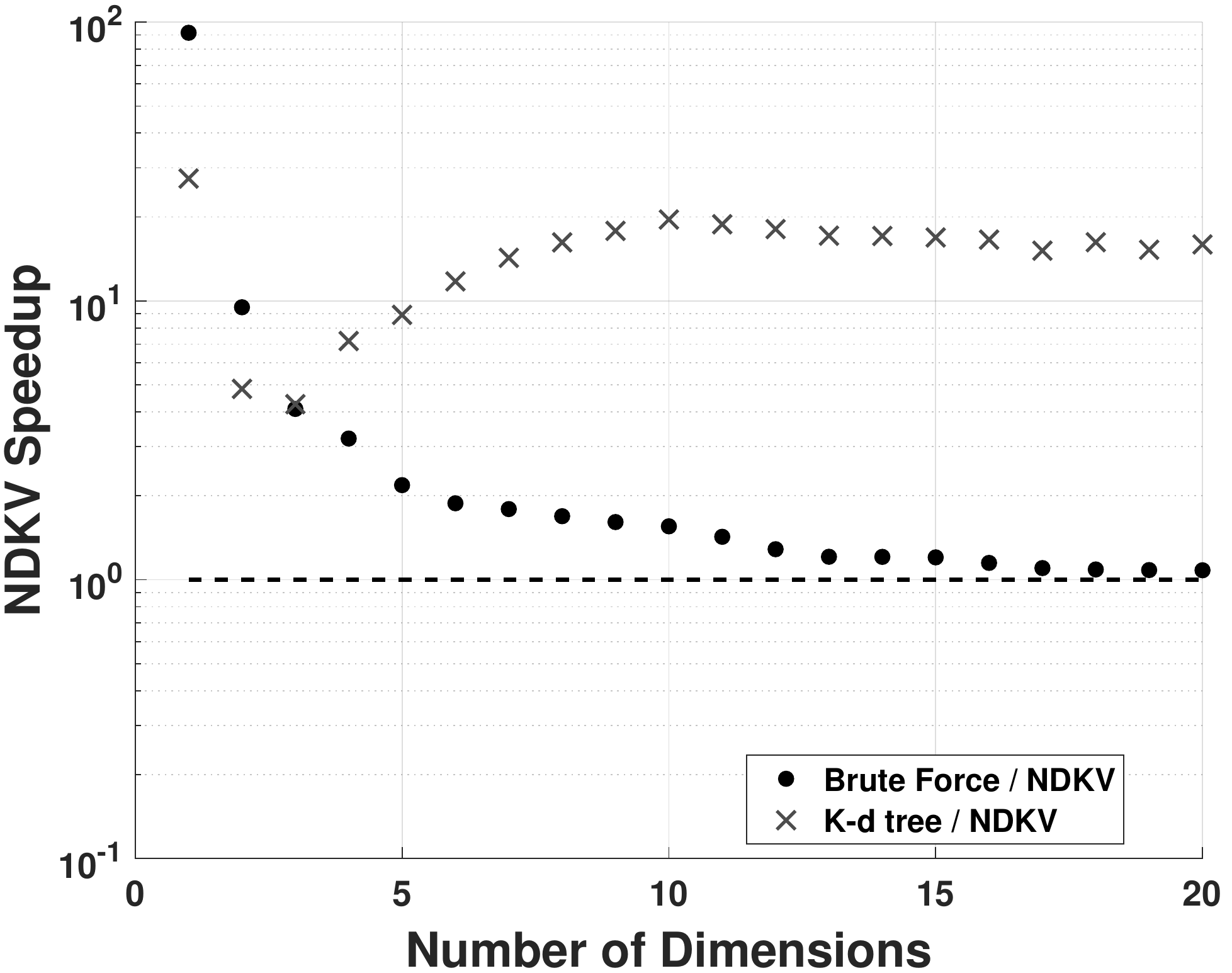}
    \caption{\Ndkv\ speedup versus number of dimensions for a database with one million elements and approximately 10\% of the database being retrieved}
    \label{fig:NumDimComp10.0}
\end{figure}

Figures \ref{fig:NumDimComp0.01} through \ref{fig:NumDimComp10.0} compare how the number of dimensions affects search time differences between \ndkv\ and the brute force and $k$-d tree algorithms. These figures were generated on a database of one million elements with a number of dimensions that ranged from one dimension to 20 dimensions. The horizontal axis of Figs. \ref{fig:NumDimComp0.01} through \ref{fig:NumDimComp10.0} shows the number of dimensions of the database and the vertical axis shows the \ndkv\ speedup. The \ndkv\ speedup is quantitatively defined as the search time of another algorithm (i.e. brute force or $k$-d tree) divided by the search time of the \ndkv\ algorithm. If this number is greater than one, the \ndkv\ is faster than the algorithm being compared against; if it is less than one, then the \ndkv\ is slower. In particular, Fig. \ref{fig:NumDimComp0.01} was created using a range search that retrieved approximately 0.01\% of the database, and Fig. \ref{fig:NumDimComp0.1} retrieved approximately 0.1\% of the database. Figure \ref{fig:NumDimComp1.0} retrieved approximately 1\% of the database, and Fig. \ref{fig:NumDimComp10.0} retrieved approximately 10\% of the database.

Figures \ref{fig:NumDimComp0.01} through \ref{fig:NumDimComp10.0} show that the \ndkv\ is faster than the other two algorithms in all instances tested except for the seven dimensional case in Fig. \ref{fig:NumDimComp0.01}. In this instance, the \ndkv\ speedup was 0.999, meaning that the $k$-d tree search time was almost identical to that of the \ndkv. In general, the \ndkv\ speedup for brute force increases as the number of dimensions decreases. The \ndkv\ speedup for $k$-d tree initially decreases as the dimensionality of the database increases, but then the \ndkv\ speedup starts increasing again starting somewhere between dimensions three and seven. Then, the \ndkv\ speedup asymptotically approaches a final value around ten. In general, the \ndkv\ speedup for brute force increases as the percentage of elements retrieved from the database decreases. The \ndkv\ speedup for $k$-d tree has the opposite trend; however, the effect that percentage of elements retrieved from the database has on the \ndkv\ speedup for $k$-d tree is not nearly as great as the effect it has on the \ndkv\ speedup for brute force.

\begin{figure}[!h]
    \centering
    \includegraphics[width=0.75\linewidth]{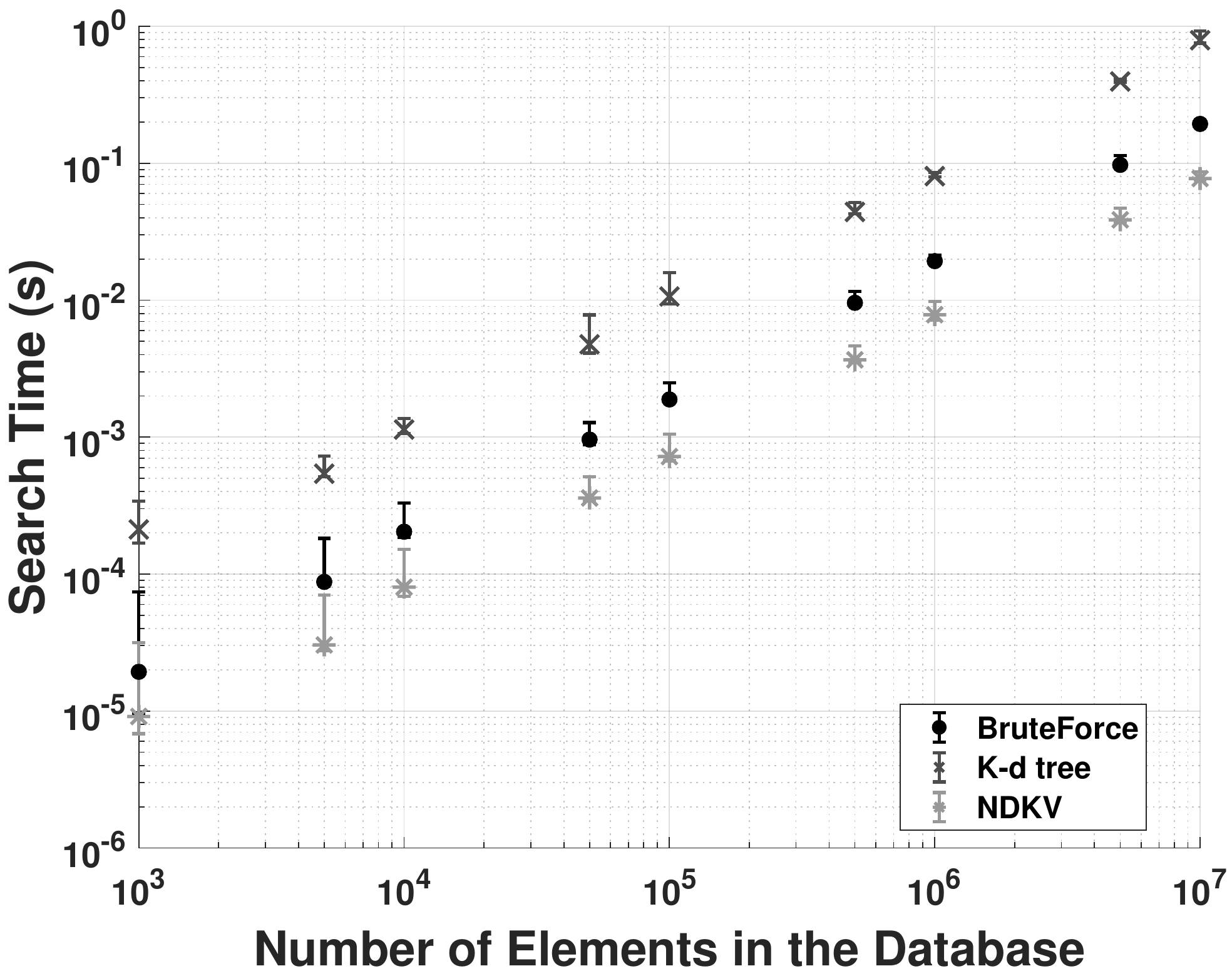}
    \caption{Search time versus number of elements for a six dimensional database with five percent of the database being retrieved}
    \label{fig:NumElComp}
\end{figure}

Figure \ref{fig:NumElComp} compares the search time of the three algorithms, \ndkv, brute force, and $k$-d tree, as a function of the number of elements in the database; the database has six dimensional elements, and the search queries were designed to return approximately five percent of the database. Each data point and associated error bar in Fig. \ref{fig:NumElComp} was created by performing the same search query 100 times for each algorithm. The mean of the 100 search queries is shown as a point, and the error bars associated with each point extend to the slowest and fastest search time for each query\footnote{The slowest and fastest search times were chosen for the error bars rather than the 95\% confidence interval of the $t$-distribution, as the 95\% confidence interval of the $t$-distribution is too small to be visible in the figure.}. 

Figure \ref{fig:NumElComp} shows that the \ndkv\ has the fastest average search time of the three algorithms for every database size; the brute force method is the second fastest, and $k$-d tree is the slowest. The difference between the average \ndkv\ search time and the brute force time is typically less than an order of magnitude, but the difference between \ndkv\ and $k$-d tree is larger than an order of magnitude in some instances. The error bars follow similar trends to the average values. 

\begin{figure}[!h]
    \centering
    \includegraphics[width=0.75\linewidth]{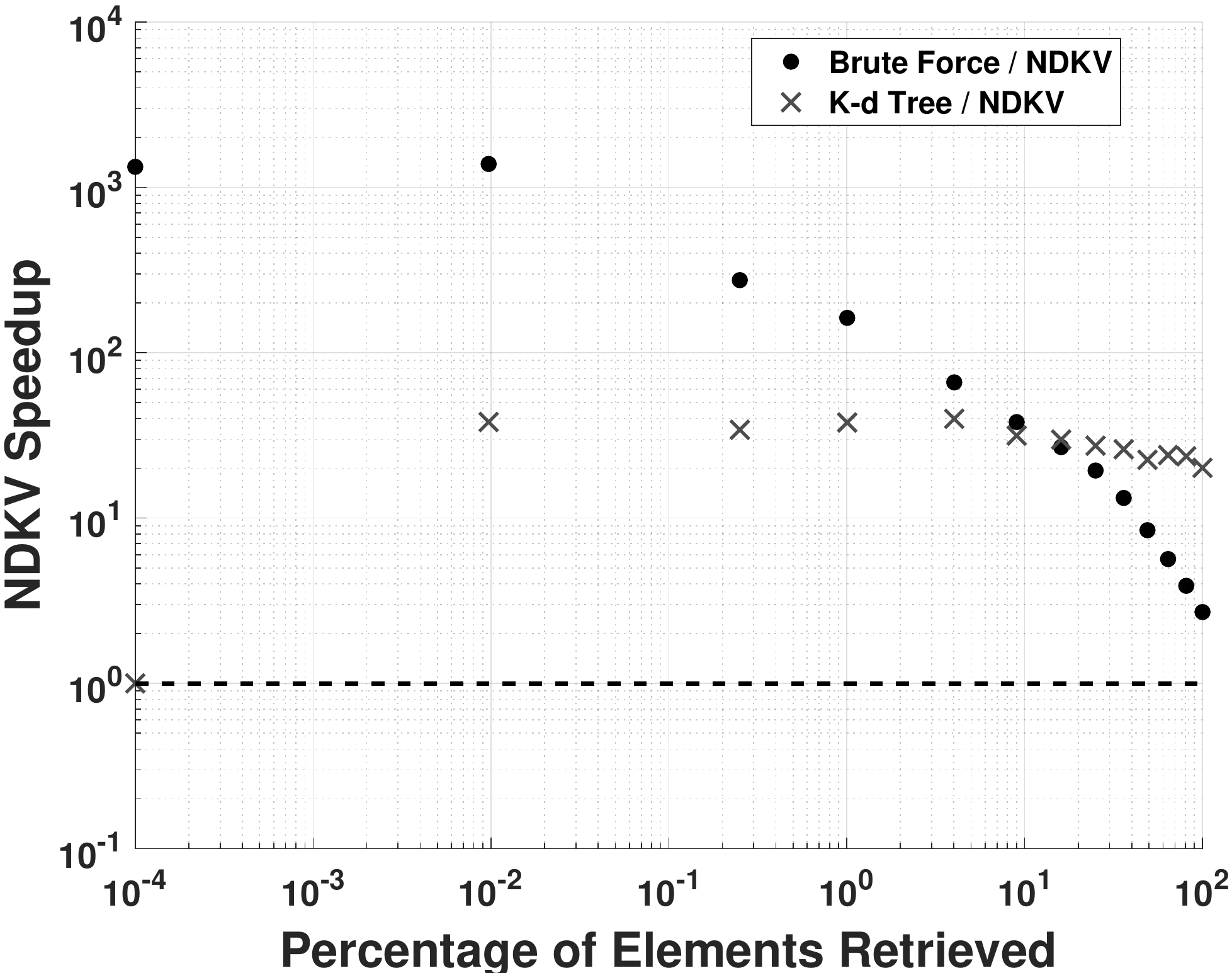}
    \caption{\Ndkv\ speedup versus percentage of the database being retrieved for a two-dimensional database with one million elements}
    \label{fig:PerComp2Dim}
\end{figure}

\begin{figure}[!h]
    \centering
    \includegraphics[width=0.75\linewidth]{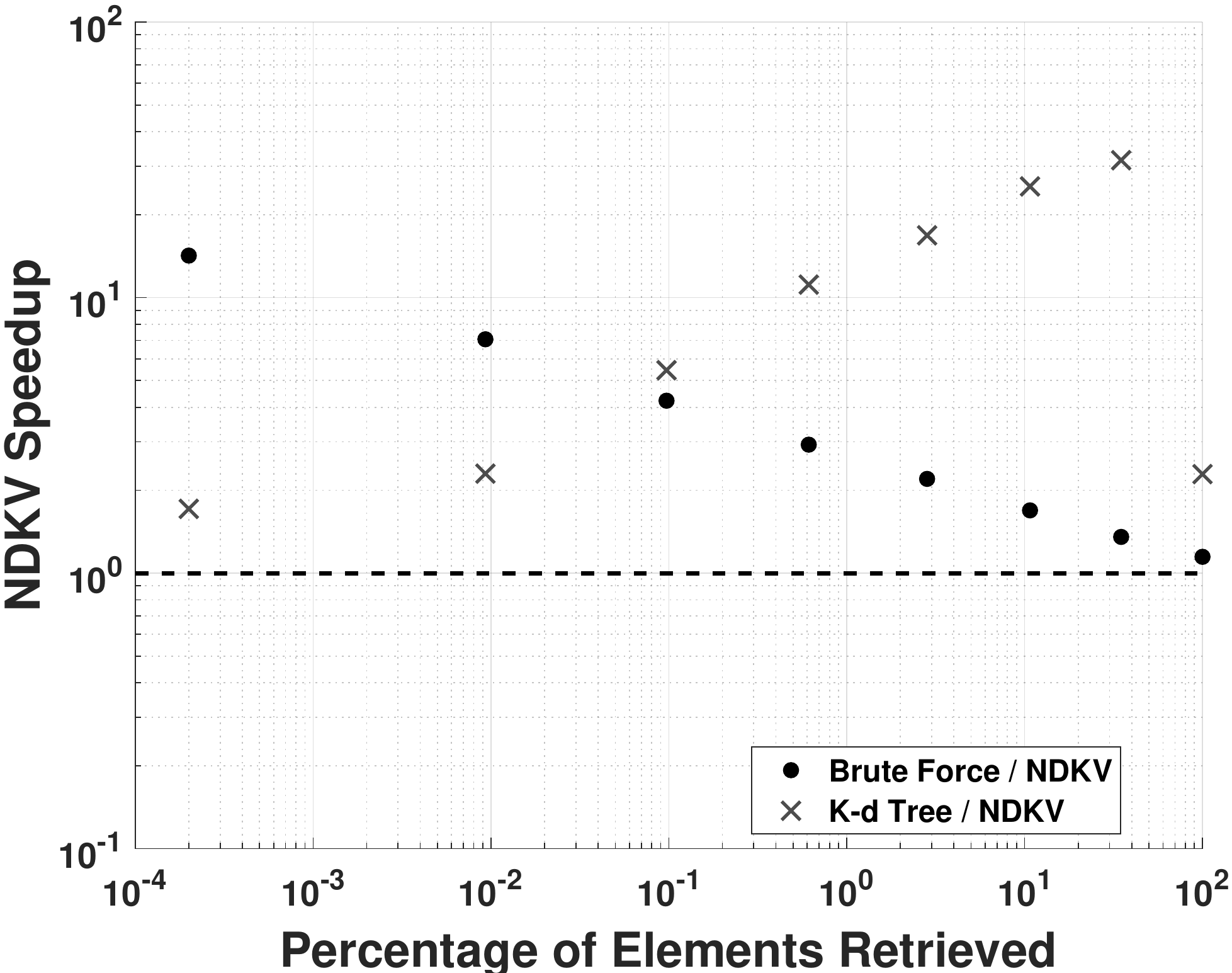}
    \caption{\Ndkv\ speedup versus percentage of the database being retrieved for a ten-dimensional database with one million elements}
    \label{fig:PerComp10Dim}
\end{figure}

Figures \ref{fig:PerComp2Dim} and \ref{fig:PerComp10Dim} compare the \ndkv\ speedup to the percentage of the database retrieved from a database of one million elements. Figure \ref{fig:PerComp2Dim} was generated using a two-dimensional database, and Fig. \ref{fig:PerComp10Dim} was generated using a ten-dimensional database. Each of the data points used to calculate the \ndkv\ speedup in these two figures was generated by measuring the search time of 1,000 random searches. 


Figures \ref{fig:PerComp2Dim} and \ref{fig:PerComp10Dim} show that \ndkv\ is faster than the other two algorithms in every case except for when $10^{-4}\%$ of the database is retrieved in Fig. \ref{fig:PerComp2Dim}. In this instance, the \ndkv\ speedup is 1.0, meaning that the \ndkv\ and $k$-d tree had the same search time. The \ndkv\ speedup for brute-force decreases in both figures as the percentage of the database that is retrieved increases. In addition, the \ndkv\ speedup for brute force in ten dimensions is less than in the two dimensional case. In the two dimensional case (see Fig.~\ref{fig:PerComp2Dim}), the \ndkv\ speedup for $k$-d tree initially increases, and then remains approximately constant as the percentage of elements retrieved increases. In the ten-dimensional database (see Fig.~\ref{fig:PerComp10Dim}), the \ndkv\ speedup for $k$-d tree increases as the percentage of elements retrieved increases, except for the final data point, 100\% of the database, where the \ndkv\ speedup decreases.

\subsection{Preprocessing performance}

\begin{figure}[!h]
    \centering
    \includegraphics[width=0.75\linewidth]{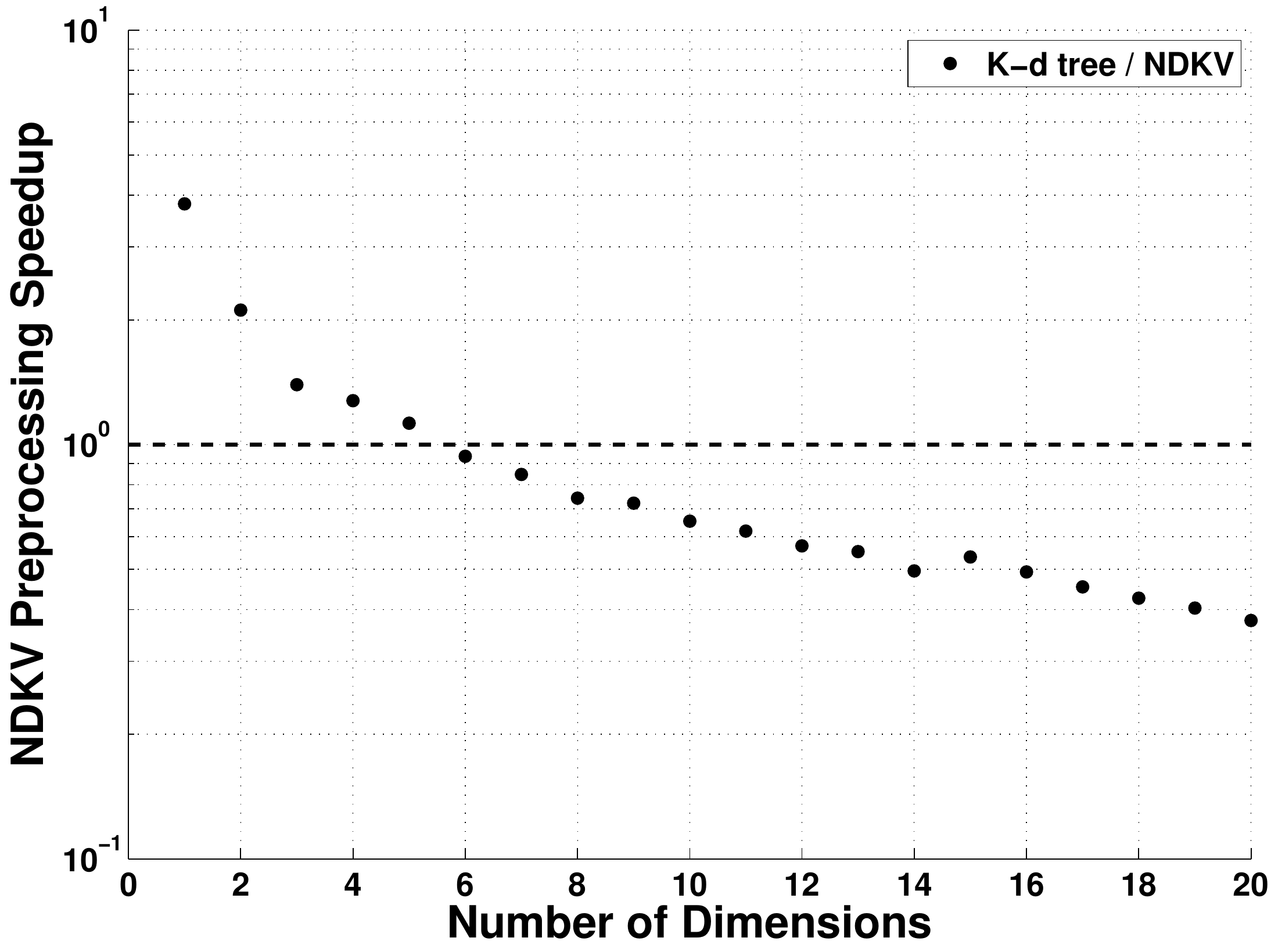}
    \caption{\Ndkv\ preprocessing speedup versus a $k$-d tree for a database with one million elements}
    \label{fig:preprocessing}
\end{figure}

In addition to the search performance of the \ndkv, the time the algorithm requires to generate the database structure and the auxiliary databases has also been studied. To that end, a direct comparison has been performed between the preprocessing times of the \ndkv\ and $k$-d tree. Figure~\ref{fig:preprocessing} shows the results of this study for a database of one million elements and a number of dimensions ranging from 1 to 20. 
As it can be seen, the preprocessing of the \ndkv\ takes less time than the $k$-d tree if the number of dimensions is low (5 dimensions or less). However, as the number of dimensions of the database increases, the \ndkv\ requires more time than the $k$-d tree to perform the preprocessing. This is due to the increased number of database sorting processes that the \ndkv\ must perform as the number of dimensions increases.

\section{The one dimensional k-vector}

Although in general the process described for the case of the \ndkv\ can be followed without any modification when dealing with one dimensional databases, due to its special particularities and properties, it is interesting to study in more detail the case of the one dimensional \kv. The \kv\ is an already known methodology that has been used in several applications~\cite{Pyramid,InvFun,David,Rogers}. However, in this section we include some important modifications compared to previous works that come as a result of the development of the \ndkv. Examples of these modifications include the database structure, the definition of the mapping function, and an alternative way to remove the extraneous elements. Additionally, a study of the algorithm complexity is also included. We study these modifications  in more detail in the following subsections.

\subsection{Preprocessing}

Dealing with one dimensional databases introduces two key differences during the preprocessing effort of the algorithm. First, the algorithm does not require generating sub-databases, because searching in multiple dimensions is no longer required. Second, the index array does not provide any useful information since the data structure is already sorted in the only dimension of the problem. This means that the preprocessing effort can be greatly simplified for the case of one dimensional databases.

Therefore, the database structure is based on sorting the whole database in ascending order, and does not require any additional processes. Then, the mapping function is defined over this sorted structure, and \kv\ and \kv\ lines arrays are generated for the whole database (no sub-database is defined). This implies that the \kv\ array is composed of $n_k$ integer numbers, while the \kv\ line array becomes just two real numbers, the slope of the mapping function $m$, and its intercept $q$.

\subsection{Searching process}

The searching process in the one dimensional \kv\ is straightforward. The searching range is first approximated as a boundary range in the \kv\ line array using Eq.~\eqref{eq:ajbj}, and then a search is performed on the extreme elements of the range to check for possible extraneous elements. As in the case of the \ndkv, this can be done by a linear search, or if the database is highly non-linear, with a binary search. This allows the algorithm to retrieve all the elements contained in the searching range without further operations.

\subsection{Algorithm complexity}

In this subsection we deal with the evaluation of the algorithm complexity of the one dimensional \kv. In that regard, two different scenarios are considered, the average case, and the worst case scenario for the algorithm.

In an average case, there are $n/n_k$ elements in a given range $[\B{z}(i),\B{z}(i+1)]$ $\forall i\in\{1,\dots,n_k-1\}$ defined by the mapping function. This means that the expected number of extraneous elements during the searching process is $n/n_k$. Then, if a linear search is performed, the algorithm will require an average of $n/n_k$ operations to finish the searching process, while a binary search applied at the \kv\ array boundaries has a complexity of $\mathcal{O}\left(\log(n/n_k)\right)$. Note that when following this methodology, it is possible to maintain the speed of the algorithm regardless of the number of elements contained in the database as long as the relation $n/n_k$ is maintained.

In the worst case scenario for the methodology, all the elements of the database except one are contained in a particular range $[\B{z}(i),\B{z}(i+1)]$ of the searching space. This implies that for a linear search would require, on average, the evaluation of $(n-1)/2$ elements from the database, which greatly deteriorates the search performance of the algorithm. For this reason, a linear search in these cases is not recommended. Instead, and since this is a very non-linear database, a binary search approach must be performed as commented in the previous subsection. That way, a complexity of $\mathcal{O}(\log n)$ is obtained where the process is equivalent to doing a pure binary search.

\section{Reducing the memory required}
\label{sec:options}

There are some applications where it is useful to reduce the amount of memory required for the algorithm. Examples include databases with a very large number of elements and dimensions, or the use of systems with limitations in memory. For this reason, in this section we discuss the possibilities of memory reduction in the \ndkv\ methodology. 

In particular, two approaches are considered in the following subsections. The first one is based on removing the necessity of the index array by fixing the dimension in which the projection is performed during the searching process. The second approach deals with the possibility of removing the \kv\ and the \kv\ line arrays by substituting them for a binary search technique. Note that performing these modifications produces a reduction in the search performance of the algorithm, but not in the algorithm complexity.

\subsection{Removing the index array}

The index array is the auxiliary database that generally requires more memory, because it is the same size as the database. Therefore, it is the first candidate to focus on in order to reduce the amount of memory required. If the index array is removed, it is no longer possible to select the most convenient dimension for the algorithm to perform a projection in, as it is no longer possible to define an effective ordering of the dimensions. Therefore, in this case the \ndkv\ focuses on just two dimensions, which correspond to the ones used in the definition of the database structure. That way, the \kv\ and \kv\ line arrays are only defined in the first dimension of the problem, and thus, they have size $n_{db}n_k$ and $2n_{db}$ respectively. 

The removal of the index array, however, leads to a new problem. Since we have removed the information about the last dimension of the database from the \kv\ array, we can no longer discard entire sub-databases during the searching process, which increases the number of elements that have to be evaluated. This difficulty can be overcome by storing the values of the smallest components in the last dimension of each sub-database in an array. Then, during the searching process, a binary search will be performed in this array to identify which sub-databases need to be searched. This modification does not change the overall algorithm complexity, because the binary search is of order $\log n_{db}$, which is much smaller than the other effects.

\subsection{Removing the \kv\ array and \kv\ line array}

It is also possible to reduce the size of the auxiliary database by removing the \kv\ array and \kv\ line array. This can be done regardless of whether the index array is removed or not. The idea of this modification is the following. Instead of using the \kv\ to locate the boundary points of the searching range in each dimension, a binary search technique is applied, using the information of the sorted database in each dimension provided by the index array or the database structure. This means that for each dimension, two binary searches of complexity $\mathcal{O}(\log n)$ must be performed that correspond to the two boundary values of the range. Once these elements are located, the process continues as described previously in the searching process.

It is important to note that although the computed speed of this modification is worse than the original methodology, its complexity is maintained. In particular, this modification has a complexity of $\mathcal{O}(d\log n + nd(k/n)^{(2/d)})$, where the second term is dominant. Therefore, the order of magnitude of the algorithm complexity is still $\mathcal{O}(nd(k/n)^{(2/d)})$, the same as in the original methodology. In fact, having removed the \kv\ from the process, the resultant algorithm is reminiscent of a modified projection method applied to two dimensions.

\subsection{Search performance of the modification}

\begin{figure}[!h]
    \centering
    \includegraphics[width=0.75\linewidth]{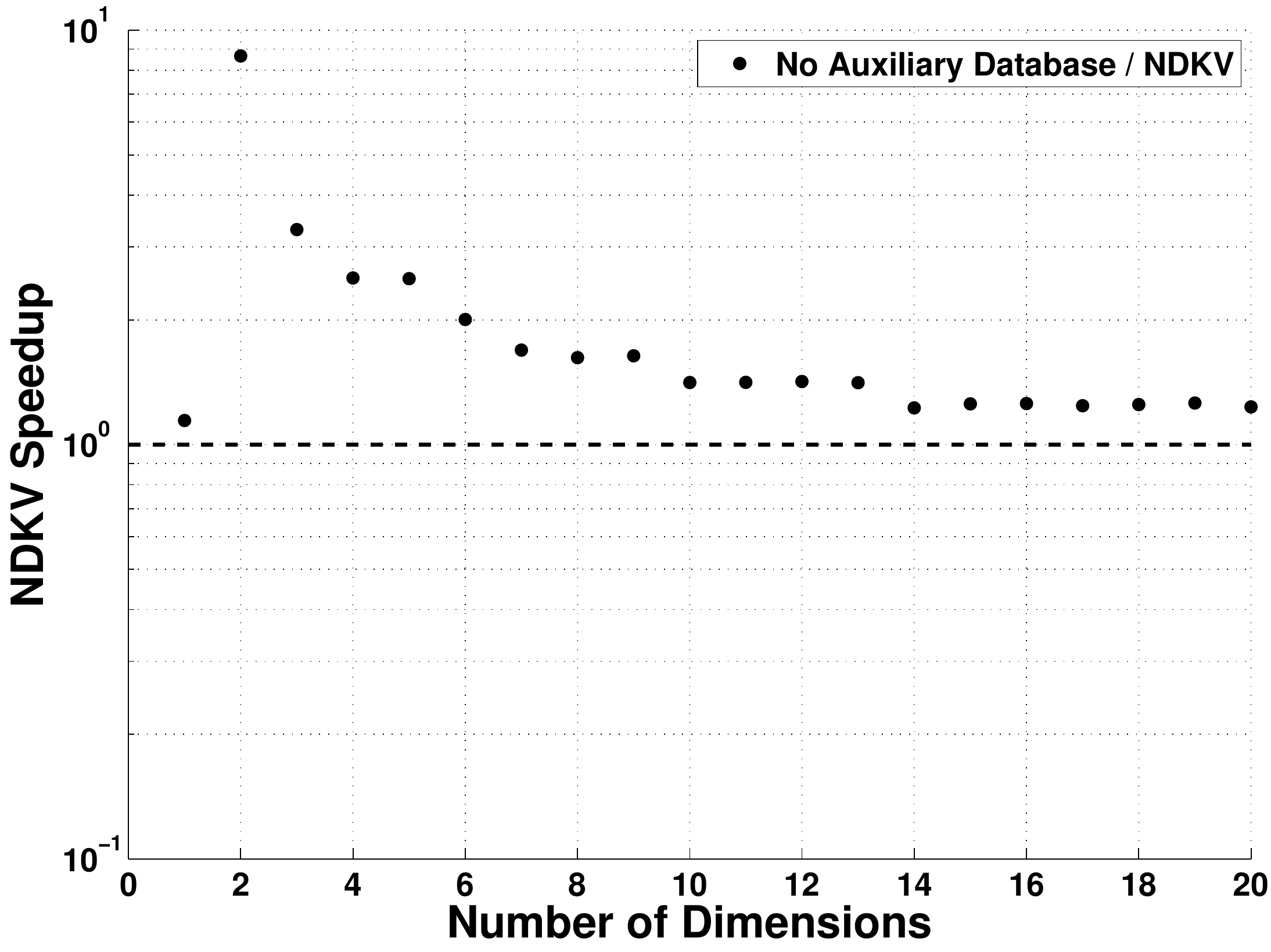}
    \caption{\Ndkv\ speedup versus algorithm not using auxiliary databases for a database with one million elements and approximately 1\% of the database being retrieved}
    \label{fig:no_auxiliary}
\end{figure}

In this subsection we compare the search performance of the \ndkv\ with its modified version as described in this section. In particular, we consider the case with no index array, no \kv\ array, and no \kv\ line array. Figure~\ref{fig:no_auxiliary} shows the results of this comparison for a database of one million points as a function of the number of dimensions, where approximately 1\% of the database was retrieved. 
As it can be seen, the original \ndkv\ outperforms the modified version regardless of the number of dimensions of the database, where the speed differences come from the advantage that the \ndkv\ has due to its auxiliary databases. The most noticeable difference is when dealing with databases in two dimensions, because the \ndkv's auxiliary databases provide a better map of the search space. This difference becomes less relevant as the number of dimensions of the database increases. Note too that this behaviour confirms the algorithm complexity study performed in this section. Moreover, when comparing Fig.~\ref{fig:no_auxiliary} with Fig.~\ref{fig:NumDimComp1.0}, we can see that this modified version of the \ndkv\ is still faster than a brute force or a $k$-d tree  approach for the cases studied.

\section{Conclusions}

This work presents the \ndkv\ algorithm, a methodology to perform orthogonal range searching in static databases with multiple dimensions. The \ndkv\ is shown to be fast, easy to implement, and adaptable to the memory requirements of the problem to be solved. In addition, performance comparisons between the \ndkv\ and other common algorithms from the literature, $k$-d tree and brute force approach, are shown. In the scenarios studied, the algorithm has proven to be faster than the aforementioned methodologies. Moreover, for all the tests performed during the creation of this article, the \ndkv\ was always faster than a brute force approach regardless of the number of elements retrieved, the size of the database, or the dimensionality of the database. These tests included the retrieval of the whole database and problems with more than one hundred dimensions.

The basic idea behind the \ndkv\ is to find the dimension where a projection can be performed such that the expected number of points to be evaluated in each dimension is minimized. To achieve that goal, an auxiliary database is generated that can obtain a good approximation of the number elements that meet the searching criteria in each dimension, which in turn can be used to determine which dimension to perform the projection in. Furthermore, the database structure of the \ndkv, its auxiliary databases, and the whole searching process can be defined completely using vectorial notation, which simplifies the formulation of the method and eases the implementation of the algorithm in the majority of systems. 

The \ndkv\ requires a one time preprocessing effort that first rearranges the database following the database structure presented, and then generates the auxiliary databases. Therefore, although not impossible, the capabilities of this methodology for dynamic databases (databases that have elements added or removed over time) are reduced, because, in general, these structures must be updated when a database modification is performed. Nevertheless, for static databases, this is not a limitation, and the algorithm is able to work with the original preprocessing indefinitely.

In addition, the \ndkv\ easily adapts to the memory available for any given problem. In that sense, several possibilities of modification for the algorithm have been presented. These modifications even included the case where no auxiliary database was defined, and thus, only the original sorted database was required. For a general application, the size of the \kv\ and \kv\ line arrays can adapt to the memory available for each particular application. That way, if more memory is available, the number of operations preformed can be reduced. This effect is due to the fact that the initial range approximation from the \kv s is closer to the actual searching range. Furthermore, if only an approximated range is required for a given application, then the \ndkv\ becomes even more efficient for two reasons. First, the algorithm no longer requires checking the extreme elements of the points retrieved by the \kv s. Second, it is possible to set the values of the approximated ranges through the mapping function. This provides perfect control over the actual searching ranges of the algorithm.

Although the obvious application of this algorithm is obtaining multidimensional elements inside any orthogonal searching range in massive databases, the algorithm can be used in other applications such as iso-surface identification, function inversion, enhancing compressive sampling, and random sampling. Moreover, based on the results obtained, the algorithm is especially useful when a large proportion of the elements from the database are retrieved. Examples of these types of applications include iso-surface identification and the computation of communications coverage in space missions. Nevertheless, in the problems where the number of elements retrieved is always very small, such as the nearest neighbor problem, the \ndkv\ has also proven to be very fast, obtaining a search performance comparable with other methodologies such as a $k$-d tree. Such applications using the \ndkv\ will be studied in future works. 

\section*{Acknowledgments}
David Arnas acknowledges the support of the Spanish Ministry of Economy and Competitiveness, Project No. ESP2017--87113--R (AEI/FEDER, UE); and the Aragon Government and European Social Fund (group E24\_17R). Carl Leake acknowledges the support of a NASA Space Technology Research Fellowship [NSTRF 2019], Grant \#: 80NSSC19K1152.

\end{document}